\documentclass{article}

     \usepackage[nonatbib,final]{neurips_2020} %camera ready

\usepackage[utf8]{inputenc} % allow utf-8 input
\usepackage[T1]{fontenc}    % use 8-bit T1 fonts
\usepackage{hyperref}       % hyperlinks
\usepackage{url}            % simple URL typesetting
\usepackage{booktabs}       % professional-quality tables
\usepackage{amsfonts}       % blackboard math symbols
\usepackage{nicefrac}       % compact symbols for 1/2, etc.
\usepackage{microtype}      % microtypography
\usepackage[square,numbers]{natbib}
\usepackage{graphicx}

\makeatletter
\newcommand{\printfnsymbol}[1]{%
  \textsuperscript{\@fnsymbol{#1}}%
}
\makeatother

\title{Retrospective Motion Correction of MR Images using Prior-Assisted Deep Learning}

\author{
  Soumick Chatterjee\thanks{Soumick Chatterjee and Alessandro Sciarra have contributed equally to this work. 
  } \\
  Department of Biomedical Magnetic Resonance, and \\
  Data and Knowledge Engineering Group, Faculty of Computer Science \\
  Otto-von-Guericke Univeristy\\
  Magdeburg, Germany \\
  \texttt{soumick.chatterjee@ovgu.de} \\
  \And
  Alessandro Sciarra\printfnsymbol{1} \\
  MedDigit, Department of Neurology,  \\
  Medical Faculty, University Hopspital, and \\
  Department of Biomedical Magnetic Resonance \\
  Otto-von-Guericke Univeristy \\
  Magdeburg, Germany \\
  \texttt{alessandro.sciarra@med.ovgu.de} \\
  \And
  Max D{\"u}nnwald \\
  MedDigit, Department of Neurology \\
  Medical Faculty, University Hopspital, and \\
  Faculty of Computer Science \\
  Otto-von-Guericke Univeristy \\
  Magdeburg, Germany \\
  \texttt{max.duennwald@med.ovgu.de} \\
  \And
  Steffen Oeltze-Jafra \\
  MedDigit, Department of Neurology \\
  Medical Faculty, University Hopspital, \\
  German Centre for Neurodegenerative Diseases, and \\
  Center for Behavioral Brain Sciences \\
  Magdeburg, Germany \\
  \texttt{steffen.oeltze-jafra@med.ovgu.de} \\
  \And
  Andreas N{\"u}rnberger \\
  Data and Knowledge Engineering Group \\
  Faculty of Computer Science \\
  Otto-von-Guericke Univeristy, and \\
  Center for Behavioral Brain Sciences \\
  Magdeburg, Germany \\
  \texttt{andreas.nuernberger@ovgu.de} \\
  \And
  Oliver Speck \\
  Department of Biomedical Magnetic Resonance \\
  Otto-von-Guericke Univeristy,  \\
  German Centre for Neurodegenerative Diseases,  \\
  Leibniz Institute for Neurobiology, and \\
  Center for Behavioral Brain Sciences \\
  Magdeburg, Germany \\
  \texttt{oliver.speck@ovgu.de} \\
}

\let\oldmaketitle\maketitle
\renewcommand{\maketitle}{\oldmaketitle\setcounter{footnote}{0}}

\begin{document}

\maketitle

\begin{abstract}
In MRI, motion artefacts are among the most common types of artefacts. They can degrade images and render them unusable for accurate diagnosis. Traditional methods, such as prospective or retrospective motion correction, have been proposed to avoid or alleviate motion artefacts.
%to obtain images without artefacts, or partially corrupted, where they meet their limitations. 
Recently, several other methods based on deep learning approaches have been proposed to solve this problem. This work proposes to enhance the performance of existing deep learning models by the inclusion of additional information present as image priors. The proposed approach has shown promising results and will be further investigated for clinical validity.
%Motion can be a big hindrance to the diagnostic accuracy of MRIs as it can introduce a significant amount of artefacts. Various prospective and retrospective motion correction strategies have been proposed over the years. 
%In recent times, numerous deep learning based retrospective motion correction strategies have been proposed.
%This work tries to enhance the performance of existing deep learning models, by trying to utilize the available image priors to improve the motion correction outcome. The proposed approach has shown promising results and will be further investigated for clinical validity. 
\end{abstract}

\section{Introduction}
%\subsection{Motion Artefacts in MRI}
One of the most common causes of artefacts in MR imaging is patient motion~\cite{Godenschweger_2016, JEZZARD2009499, doi:10.1002/jmri.24850}.
Physiological motion, such as cardiac or respiratory movement, can be controlled by gating or by specific sequence design~\cite{doi:10.1002/mrm.22306}.
The focus of this work is on correcting for less predictable voluntary or involuntary movement of the patient, often called, bulk motion~\cite{chenevert1991effect}, which e.g. in Parkinsonism and leads to ghosting and blurring~\cite{JEZZARD2009499}.
If the level of artefacts is too high, the scan must be repeated. Even a moderate level may lead to an incomplete or inaccurate diagnosis~\cite{Budrys_2018}.

%\subsection{Current Strategies}
%The most common strategies for handling motion artefacts are retrospective motion correction (RMC) and prospective motion correction (PMC)~\cite{doi:10.1002/jmri.24850}.
The most common strategies for handling motion artefacts can be divided into three main groups: motion prevention, artefact reduction, and motion correction. 
Motion prevention spans from training the subject, foam restraints, feed and wrap (for babies), sedation till breath-hold. Artefact reduction can be achieved with faster imaging, triggering and gating, phase reordering, and similar techniques. Finally, there are motion correction techniques such as retrospective motion correction (RMC) and prospective motion correction (PMC)~\cite{doi:10.1002/jmri.24850}.
% ------------------------------------------------
% Common motion artefact mitigation strategies
% ------------------------------------------------
% Motion prevention|Artefact reduction|Motion correction
% ------------------------------------------------
% Training|Faster imaging|Navigators
% ------------------------------------------------
% Distraction|Insensitive sequences|Self-navigated trajectories (PROPELLER and alike)
% ------------------------------------------------
% Feed and wrap (for babies)|Gradient moment nulling| -----	
% ------------------------------------------------
% Foam restraints|Saturation bands|Prospective correction
% ------------------------------------------------
% Sedation|Triggering and gating|Retrospective correction
% ------------------------------------------------
% Bitebars/head holders|Phase reordering|-----	
% ------------------------------------------------
% Breathhold| ------ | -----
% ------------------------------------------------
In practice, RMC modifies the MR image data during the reconstruction process, while PMC performs a real-time adaptive update of the data acquisition with the possibility of using different tracking modalities~\cite{Godenschweger_2016}.
Recently, an increasing number of attempts using deep learning techniques have shown that it is possible to remove or reduce motion artefacts~\cite{duffy2018retrospective,jiang2019respiratory,kustner2019retrospective,pawar2018motion, pawar2019suppressing,salehi2018real,usman2020retrospective}. It has also been observed that supplying additional prior knowledge, such as image priors, can help improve the performance of deep learning based reconstructions~\cite{liang2016incorporating,souza2020enhanced}.
This work aims to remove motion artefacts from corrupted brain images by improving existing deep learning models, by exploiting available non-corrupted image priors.
%\subsection{Our proposal}
%We propose to use two different architectures. The first one is a dual-branch network where the first branch is a classical UNet~\cite{ronneberger2015unet} or ResNet~\cite{chatterjeemri2019}, and the second branch is constituted of the encoding  

\section{Methodology}
\subsection{Data Preparation}
T1, T2, and PD images of 100 subjects (for each - training, testing, and validation) from the publicly available IXI Dataset\footnote{Dataset available at: \url{https://brain-development.org/ixi-dataset/}} were used in this study. T2-weighted images were artificially corrupted with motion using a modified version of the RandomMotion transformation of TorchIO~\cite{perez-garcia_torchio_2020} (v0.17.45). The initial phase of the experiments were performed with 10 simulated movements with a rotation in the range of -1.75 to +1.75 degrees without any translation. In this modified version of the RandomMotion function, in-plane motion corruption was randomly performed in X or Y direction. 

\subsection{Image Priors}
Supplying additional images as prior knowledge along with the corrupted image can help improve the performance of deep learning models~\cite{liang2016incorporating,souza2020enhanced}. In this research, experiments were performed using two different types of image priors - similar slices from different subjects of the same MRI contrast and different MRI contrasts of the same subject. 

\paragraph{Similar Slices:}
During the motion correction, ten similar (same slice position) slices of the same MRI contrast were randomly chosen from different subjects and supplied as prior along with the motion corrupted image. The motivation behind this type of prior is that while performing motion correction on certain image, images of the same contrast but of different subjects, which are not corrupted by motion, can readily be available. In these experiments, only T2-weighted images from the IXI Dataset were used.

\paragraph{Different Contrasts of the Same Subject:}
Multiple contrasts of the same subject are usually acquired during clinical routine. If one of those various contrasts is corrupted by motion, then that image can be corrected by using the other contrasts of the same subject as prior. All three available contrasts of the IXI Dataset were co-registered against the T2-weighted images. T2-weighted images were artificially corrupted; T1 and PD images were used as priors during the correction process. 

\subsection{Network Architectures}
UNet~\cite{ronneberger2015unet} and a modified version of the ResNet~\cite{chatterjeemri2019} were used as the baselines for this work. The baseline networks were modified such that they can receive priors. Two different methods of supplying priors were tested.

\paragraph{Multi-Channel Network:}
Each motion corrupted image and the priors were concatenated on the channel dimension and were supplied to the network as multi-channel input. The baselines were receiving only one channel image as input, where for the multi-channel approach, the models received $1+n\_{prior}$ channel images as input. 

\paragraph{Dual-Branch Network:}
For this approach, modified versions of the baselines were created by adding an additional branch to the baselines for the priors. The main branch was supplied with the motion corrupted image and the priors were supplied to the auxiliary branch. For the UNet, the auxiliary branch was identical to the contraction path and the latent space of the UNet except for the number of input channels. The skip-connections were supplied only from the main branch of the network, and no skip-connections were supplied from the auxiliary branch. For the ResNet, the auxiliary branch was identical to the downsampling blocks of the ResNet up to the residual blocks, with the only difference being the number of input channels. For both network models, the main branch and the auxiliary branch generated two different latent space representations. These latent representations were combined and forwarded for generating the final output. Two different methods were considered for combining the latent spaces - by  simple addition or by concatenation and convolution with a kernel size of one - to generate the final combined latent space. This combined latent space was forwarded to the expansion path of the UNet. In the case of ResNet, this latent representation was forwarded to the residual blocks for further operation. 

\section{Results and Discussions}
\begin{figure}[!h]
\centering
\includegraphics[width=0.9\textwidth]{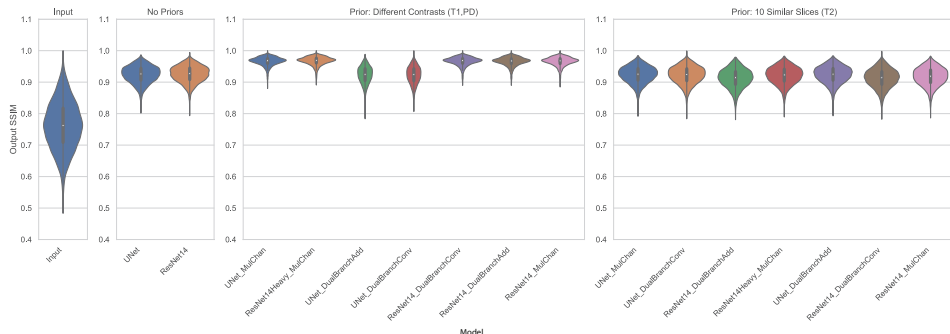}
\vspace*{-4mm}
\caption{Plots showing the performance of the various methods, based on SSIM}        \label{fig:resultsplot}
\end{figure}

\begin{figure}[!h]
\centering
\vspace*{-4mm}
\includegraphics[width=0.6\textwidth]{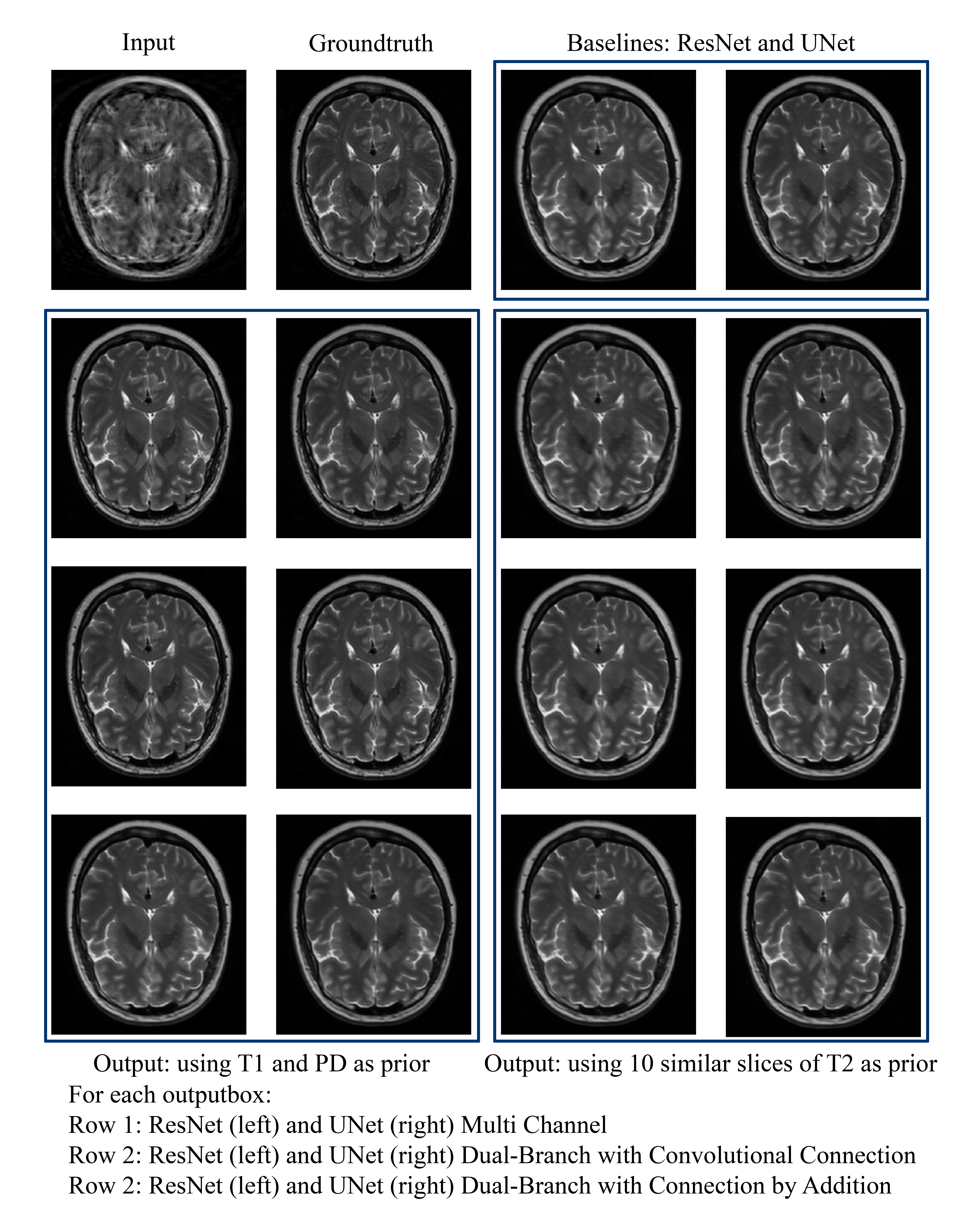}
\vspace*{-4mm}
\caption{One example slice to show the motion correction performance of the various methods}        \label{fig:image}
\end{figure}

Figure~\ref{fig:resultsplot} shows the performance of the different methods based on the SSIM~\cite{SSIM} values and Figure~\ref{fig:image} shows a representative example result. Between the two different types of priors, it was observed that supplying ten similar slices of the same contrast but of different subjects did not improve the motion correction. Nonetheless, supplying different contrasts of the same subject improved the motion correction significantly for most of the tests carried out.
%for most of the methods. 
ResNet's performance improved for both types of prior supply methods - multi-channel and dual-branch. For UNet however, only the multi-channel approach has shown significant improvement. 

\section{Conclusion and Future Work}
The initial experiments presented here have shown promising results for supplying other contrasts as prior for deep learning based motion correction. For ResNet, the multi-channel and the dual-network approach both have shown promising results. However, for UNet only the multi-channel approach has shown improvements over the baseline. The reason for the failure could be attributed to the lack of skip connections from the auxiliary branch, but the skip connections will be making it similar to the multi-channel approach. Further investigations will be performed to determine the cause of the failure, and also other strategies for supplying the priors to the UNet. Further experiments will be performed to check the performance of these approaches when the corrupted volumes and the prior volumes are not co-registered. Supplying ten similar slices as prior did not show any improvement for any of the networks. Further investigations will be performed to determine the reasons and possible solutions.  Furthermore, the strategies will be validated with clinical data. It will be assessed whether pathologies that can only be seen in a corrupted contrast, are preserved after performing the correction.

\section*{Broader Impact}
The current stage of the work has been carried out on simulated motion artefacts only and using a publicly available dataset and has been carried out in compliance with ethical standards. To the best of the authors' knowledge, there are not evident direct or indirect effects on societal aspects.
It is well known that low-quality images invalidate the clinical diagnosis. In addition, the repetition of scans involves the use of energy, money and human resources.
Therefore, researchers and clinicians working in the field of MRI may benefit from this work.
%This work does not put anyone at a disadvantage.
%There are no consequences of a possible failure of the system as at  
As this research is in its preliminary stage, it has still to be investigated whether the usage of priors exclude pathologies from the corrected data.

%At this stage of development, our work is not commercially or clinical available, hence there are not consequences of possible failure of the system.  
%Authors are required to include a statement of the broader impact of their work, including its ethical aspects and future societal consequences. 
%Authors should discuss both positive and negative outcomes, if any. For instance, authors should discuss a)  who may benefit from this research, b) who may be put at disadvantage from this research, c) what are the consequences of failure of the system, and d) whether the task/method leverages biases in the data. If authors believe this is not applicable to them, authors can simply state this.
%Use unnumbered first level headings for this section, which should go at the end of the paper. {\bf Note that this section does not count towards the eight pages of content that are allowed.}

\begin{ack}
This work was partially conducted within the context of the International Graduate School MEMoRIAL at Otto von Guericke University (OVGU) Magdeburg, Germany, kindly supported by the European Structural and Investment Funds (ESF) under the programme "Sachsen-Anhalt WISSENSCHAFT Internationalisierung" (project no. ZS/2016/08/80646). This work was also partially conducted within the context of the Initial Training Network program, HiMR, funded by the FP7 Marie Curie Actions of the European Commission, grant number FP7-PEOPLE-2012-ITN-316716, and supported by the NIH grant number 1R01-DA021146, and by the federal state of Saxony-Anhalt under grant number 'I 88'.

\end{ack}

\medskip

\small

\bibliography{mybib}

\end{document}